\documentclass{jnmp}

\usepackage{amsmath}
\JNMPnumberwithin{equation}{section}

\newtheorem{theorem}{Theorem}
\newtheorem{proposition}{Proposition}
\newtheorem{lemma}{Lemma}

\theoremstyle{definition}
\newtheorem{remark}{Remark}

\begin{document}
\renewcommand{\evenhead}{K Kajiwara and K Kimura}
\renewcommand{\oddhead}{$q$-Painlev\'e III Equation}

\thispagestyle{empty}

\FirstPageHead{10}{1}{2003}{\pageref{kajiwara-firstpage}--\pageref{kajiwara-lastpage}}{Article}

\copyrightnote{2003}{K Kajiwara and K Kimura}

\Name{On a $\boldsymbol{q}$-Difference Painlev\'e III Equation: I.\\
 Derivation, Symmetry and Riccati Type Solutions}
\label{kajiwara-firstpage}

\Author{Kenji KAJIWARA~$^\dag$ and Kinji KIMURA~$^\ddag$}

\Address{$^\dag$~Graduate School of Mathematics, Kyushu
University,\\
~~6-10-1 Hakozaki, Higashi-ku, Fukuoka 812-8512, Japan\\
~~E-mail: kaji@math.kyushu-u.ac.jp\\[10pt]
$^\ddag$~Department of Mathematics, Kobe University, 1-1 Rokko, Kobe 657-8501, Japan\\
~~E-mail: kimura@math.kobe-u.ac.jp}

\Date{Received May 10, 2002; Revised July 25, 2002; 
Accepted July 31, 2002}

\begin{abstract}
\noindent
A $q$-difference analogue of the Painlev\'e III equation is considered.
Its derivations, affine Weyl group symmetry, and two kinds of special function type
 solutions are discussed.  
\end{abstract}

\section{Introduction}
Importance of the Painlev\'e equations is widely recognized in
mathematics and mathematical physics, since the solutions plays a role
of nonlinear version of special functions. So much studies for the
Painlev\'e equations have been done from various points of view, such as
classical solutions, asymptotics, geometric or algebraic structures and
so on~\cite{LW:Painleve,Co:Painleve}.  On the other hand, discrete
analogue of the Painlev\'e equations have been studied extensively after
discovery of the singularity confinement property~\cite{SC}, which is a
discrete analogue of the Painlev\'e property.  What matters in 
studies of the discrete Painlev\'e equations was that they were in some
sense too rich and looked almost uncontrollable.  For example, number of
known discrete Painlev\'e equations is several ten's, while that for
the Painlev\'e equations is only six.

The situation has been changed after Sakai's theory~\cite{Sak}. Sakai
constructed ``space of initial conditions'', or defining manifold of a
discrete Painlev\'e equation called $q$-P$_{\rm VI}$ by blowing up
$\mathbb{P}^2$ at nine points. Based on this fact, Sakai gave a
classification of discrete Painlev\'e equations with the aid of theory
of rational surfaces, and discussed the symmetries.  The Painlev\'e
equations appear naturally from the degeneration of defining manifold
which is also interpreted as continuous limit. This theory gives us clear
view to both discrete and continuous Painlev\'e equations. It also
suggests the list of discrete Painlev\'e equations to be studied
further. 

In this paper, we shall work with the following $q$-difference equation,
\begin{gather}
 \overline{f_1}=\frac{q^{2N+1}c^2}{f_0f_1}
\, \frac{1+a_0q^nf_0}{a_0q^n+f_0},\nonumber\\
 \underline{f_0}=\frac{q^{2N+1}c^2}{f_0f_1}
\, \frac{a_1q^{-n+\nu}+f_1}{1+a_1q^{-n+\nu}f_1},\label{qP3}
\end{gather}
where $f_i=f_i(n;\nu,N)$ $(i=0,1)$ are dependent variables,
$n\in\mathbb{Z}$ is the independent variable, $\nu, N\in\mathbb{Z}$ are
parameters, and $q$, $a_0$, $a_1$, $c$ are constants. Moreover,
$\overline{f_i}$ and $\underline{f_i}$ denote $f_i(n+1;\nu,N)$ and
$f_i(n-1;\nu,N)$, respectively.  We call equation~(\ref{qP3}) a $q$-difference
Painlev\'e III equation($q$-P$_{\rm III}$).  Since $q$-P$_{\rm III}$
appears in ``Mul.6'' in Sakai's classification, it is expected that
$q$-P$_{\rm III}$ is a ``good'' equation.

The purpose of our work is to establish and to give explicit description
for the following properties of $q$-P$_{\rm III}$~(\ref{qP3}):
\begin{itemize}\vspace{-2mm}
\itemsep0mm
 \item $q$-P$_{\rm III}$ admits a continuous limit to the Painlev\'e III
       equation~(P$_{\rm III}$),
 \begin{equation}
\label{p3}
  \frac{d^2v}{dx^2}=\frac{1}{v}\left(\frac{dv}{dx}\right)^2 - \frac{1}{x}\frac{dv}{dx}
+\frac{1}{x}\left(\alpha v^2+\beta \right) 
+ \gamma v^3+ \frac{\delta}{v}, \qquad \gamma=-\delta=4,
\end{equation}
where $\alpha$, $\beta$ are parameters.
\item $q$-P$_{\rm III}$ is derived from the discrete-time Relativistic
       Toda lattice equation~\cite{Suris:dRT}, while P$_{\rm III}$ is
       derived from the relativistic Toda lattice
       equation~\cite{Ruijsenaars}.
 \item $q$-P$_{\rm III}$ can be regarded as a discrete dynamical system
       on the root lattice of type $A_2\times A_1$~\cite{Sak,KNY:qp4}. As
       a consequence, it admits symmetry of (extended) affine Weyl group
       of type $A_1^{(1)}\times A_1^{(1)}$ as the group of B\"acklund
       transformations.
 \item $q$-P$_{\rm III}$ admits two classes of classical solutions,
       namely, the solutions which are reducible to discrete Riccati
       equation, and rational solutions. There are two kinds
       of Riccati type solutions, one of which is expressed by the
       Jackson's (modified) $q$-Bessel functions, while the Riccati type
       solution of P$_{\rm III}$ is expressed by the
       (modified) Bessel functions~\cite{Okamoto:p3}.
 \item The rational solutions of P$_{\rm III}$ (\ref{p3}) are expressed as
       ratio of some special polynomials (Umemura polynomials). They
       admit determinant formula of Jacobi--Trudi type whose entries are
       given by the Laguerre polynomials~\cite{KM:p3}. The rational solutions of
       $q$-P$_{\rm III}$ are characterized by similar
       special polynomials which also admit determinant formula.\vspace{-2mm}
\end{itemize}

In this paper, being the first half of our work, we discuss the derivation,
symmetry, and particular solutions of Riccati type. Rational solutions
will be discussed in the next paper.

This paper is organized as follows. In Section~2 we derive P$_{\rm
III}$ and $q$-P$_{\rm III}$ from the~rela\-tivistic Toda lattice and
discrete-time relativistic Toda lattice, respectively. Moreover, we also
derive $q$-P$_{\rm III}$ from the birational representation of affine
Weyl group of type $A_2^{(1)}\times A_1^{(1)}$ which was presented 
in~\cite{KNY:qp4}. In Section~3, we construct two kinds of particular
solutions of Riccati type, one of which is expressed by Jackson's
modified $q$-Bessel functions. We also construct determinant formula for
these solutions.

\begin{remark}
1. Originally $q$-P$_{\rm III}$ (\ref{qP3}) was derived in \cite{KTGR} as
``asymmetric generalization'' of $q$-difference Painlev\'e II
 equation ($q$-P$_{\rm II}$)~\cite{RG:coales},
\begin{equation}
\label{qP2}
 x_{n+1}x_nx_{n-1}=\xi \frac{x_n+\zeta \lambda^{n}}{1+\zeta \lambda^n x_n},
\end{equation}
where $\zeta$, $\xi$ are parameters, and $\lambda$ is a
constant. In fact, putting
\begin{gather}
q=\lambda^{-2},\qquad \nu=0,\qquad a_1 = \frac{1}{a_0\lambda},\nonumber\\
x_{2n}=f_1(n;0,N),\qquad x_{2n+1}=f_0(n;0,N),\label{qP2:specialize}
\end{gather}
$q$-P$_{\rm III}$ (\ref{qP3}) is reduced to $q$-P$_{\rm II}$ (\ref{qP2}) with
\begin{equation}
 \xi=\lambda^{-4N+2}c^2,\qquad \zeta=\frac{1}{a_0\lambda}.
\end{equation} 

2. The following $q$-difference equation,
\begin{equation}\label{qp3:GR}
 w_{n+1}w_{n-1}=\frac{cd(w_n-aq^n)(w_n-bq^n)}{(w_n-c)(w_n-d)},
\end{equation}
is also identified as a
$q$-difference Painlev\'e III
equation~\cite{RG:coales,RGH:dP}. Here, $q$ is a constant and $a$, $b$,
$c$, $d$ are parameters. A class of particular solutions in
terms of Jackson's $q$-Bessel functions is constructed
 in~\cite{KOS:dP3}.  It looks that equation~(\ref{qp3:GR}) has a different nature
compared to our $q$-P$_{\rm III}$~(\ref{qP3}), and it may be natural to
regard equation~(\ref{qp3:GR}) as a degenerate case of the $q$-difference
Painlev\'e VI equation ($q$-P$_{\rm VI}$)~\cite{JS:qp6,Sak},
\begin{gather}
 y_ny_{n+1}=\frac{a_3a_4(z_{n+1}-b_1q^n)(z_{n+1}-b_2q^n)}
{(z_{n+1}-b_3)(z_{n+1}-b_4)},\nonumber\\
 z_nz_{n+1}=\frac{b_3b_4(y_{n}-a_1q^n)(y_{n}-a_2q^n)}
{(y_{n}-a_3)(y_{n}-a_4)},\qquad
\frac{b_1b_2}{b_3b_4}=q\frac{a_1a_2}{a_3a_4},
\end{gather}
where $a_i$ and $b_i$ ($i=1,2,3,4$) are parameters. 
\end{remark}

\section{Derivations of $\boldsymbol{q}$-P$\boldsymbol{{}_{\rm III}}$}
\subsection{Derivation of P$\boldsymbol{{}_{\rm III}}$ 
from relativistic Toda lattice}

In this section we derive P$_{\rm III}$ (\ref{p3}) from the relativistic Toda
lattice equation~\cite{Ruijsenaars,Suris:dRT},
\begin{gather}
  \frac{d}{dt}d_n=d_n(c_n-c_{n-1}), \nonumber\\
  \frac{d}{dt}c_n=c_n(d_{n+1}+c_{n+1}-d_n-c_{n-1}),
\qquad n\in\mathbb{Z}.
\end{gather}
We introduce the variables $V_n$ and $K_n$ by
\begin{equation}
c_n = -\frac{V_nK_n}{K_{n-1}},\qquad
d_n = -K_{n-1}+\frac{V_{n-1}K_{n-1}}{K_{n-2}},
\end{equation}
which yields 
\begin{gather}
 \frac{d}{dt}\log V_n = (K_{n-1}+V_{n-1}) - (K_n+V_n),\nonumber\\
 \frac{d}{dt}K_n = K_nV_n - K_{n+1}V_{n+1},
\qquad n\in\mathbb{Z}.\label{rt:KV}
\end{gather}
We next impose two-periodicity on equation~(\ref{rt:KV}). Then we have,
\begin{gather}
\frac{d}{dt}\log V_0 = (K_1+V_1) - (K_0+V_0),\nonumber\\
\frac{d}{dt}\log V_1 = (K_0+V_0) - (K_1+V_1),\nonumber\\
\frac{d}{dt}K_0 = K_0V_0 - K_{1}V_{1},\nonumber\\
\frac{d}{dt}K_1 = K_1V_1 - K_{0}V_{0}.\label{rt:KV2p}
\end{gather}
In order to derive P$_{\rm III}$ (\ref{p3}), we introduce additional
constants $\alpha_n$ $(n=0,1)$ in such a way that
\begin{gather}
\frac{d}{dt}\log V_0 = (K_1+V_1) - (K_0+V_0)+\alpha_0,\nonumber\\
\frac{d}{dt}\log V_1 = (K_0+V_0) - (K_1+V_1)+\alpha_1,\nonumber\\
\frac{d}{dt}K_0 = K_0V_0 - K_{1}V_{1},\nonumber\\
\frac{d}{dt}K_1 = K_1V_1 - K_{0}V_{0}.\label{rt3:2p}
\end{gather}
We note that integrability is still kept under this generalization in a
sense that it admits the Lax pair.  
It is easy to see from equation~(\ref{rt3:2p}) that we have $K_0+K_1={\rm
const}$ and $\frac{d}{dt}\log(V_0V_1)=\alpha_0+\alpha_1$.
We normalize them so that
\begin{equation}
\alpha_0+\alpha_1=1,\qquad V_0V_1 = {\rm e}^t,\qquad K_0 + K_1 = \beta_0.
\label{rt3:2p:constraint}
\end{equation}
Under this normalization, one can write down the
equations for $V_0$ and $K_1$. Eliminating $K_1$, we obtain
\begin{equation}
\label{p3:exp}
 \frac{d^2V_0}{dt^2}=\frac{1}{V_0}\left(\frac{dV_0}{dt}\right)^2
+V_0^3-(\alpha_0+\beta_0) V_0^2 + (\beta_0-\alpha_0+1){\rm e}^t 
- \frac{{\rm e}^{2t}}{V_0},
\end{equation}
or 
\begin{equation}
  \frac{d^2V_0}{ds^2}=\frac{1}{V_0}\left(\frac{dV_0}{ds}\right)^2
 - \frac{1}{s}\frac{dV_0}{ds}
+\frac{V_0^2}{s^2}\left(V_0 - (\alpha_0+\beta_0) \right) 
+ \frac{\beta_0-\alpha_0+1}{s}- \frac{1}{V_0},\label{p3'}
\end{equation}
where $s={\rm e}^t$, which is a version of P$_{\rm III}$
(named as ``P$_{\rm III'}$'' in~\cite{Okamoto:p3}). Equation~(\ref{p3'})
is rewritten into the ordinary form of P$_{\rm III}$~(\ref{p3})
with $\alpha=-4(\alpha_0+\beta_0)$, $\beta=4(-\alpha_0+\beta_0+1)$,
$\gamma=-\delta=4$ by putting $s=x^2$ and $V_0=xv$.

\begin{remark}
Without introducing additional constants $\alpha_n$, we obtain the
equation for $V_0$ as
 \begin{equation}
\label{p12}
 \frac{d^2V_0}{dt^2}=\frac{1}{V_0}\left(\frac{dV_0}{dt}\right)^2
+V_0^3-\beta_0 V_0^2 + \beta_0- \frac{1}{V_0}.
\end{equation}
Equation~(\ref{p12}) is classified as ``XII'' in Gambier's
classification~\cite{Ince}, and integrated in terms of the elliptic
functions. Therefore, such generalization of the relativistic Toda
lattice equation is crucial in order to derive P$_{\rm III}$. 
Such situation can be seen for other Painlev\'e equations, where
they are derived by the similar generalization of proper soliton
equations~\cite{Adler,NY:p4,NY:higher,NRGO}.
\end{remark}

\subsection{Derivation of $\boldsymbol{q}$-P$\boldsymbol{{}_{\rm III}}$ 
from discrete-time relativistic Toda lattice}

Let us next consider the discrete case. We start from the discrete-time
relativistic Toda lattice equation~\cite{Suris:dRT} in the following
form,
\begin{gather}
 \frac{\overline{d}_n}{d_n}=\frac{a_{n-1}-d_{n-1}}{a_n-d_n},\qquad
 \frac{\overline{c}_n}{c_n}=\frac{a_{n-1}+c_{n-1}}{a_n+c_n},\nonumber\\
  a_n = 1+d_n+\dfrac{c_{n+1}}{a_{n+1}},\qquad
n\in\mathbb{Z},\label{dRT:cd}
\end{gather}
where $a_n=a_n(t)$, $c_n=c_n(t)$, $d_n=d_n(t)$ and
$\overline{\phantom{d}}$ denotes the value at $t+1$. We note that for
convenience we have inverted the space direction $n$ of the original
equation in~\cite{Suris:dRT}. We introduce the variables $K_n=K_n(t)$,
$V_n=V_n(t)$ by
\begin{equation}
 \frac{c_n}{a_n}=\overline{V}_n,\qquad a_n=K_n.\label{cd2KV}
\end{equation}
Then equation~(\ref{dRT:cd}) is rewritten as
\begin{gather}
  \frac{\overline{V}_n}{V_n}=\frac{K_{n-1}+V_{n-1}}{K_n+V_n}, \nonumber\\
  \frac{\overline{K}_{n+1}}{K_n}=\frac{1+\overline{V}_n}{1+\overline{V}_{n+1}}
\, \frac{\overline{K}_{n+1}+\overline{V}_{n+1}}{\overline{K}_n+\overline{V}_n}.
\label{dRT:KV}
\end{gather}
We next impose two-periodicity on equation~(\ref{dRT:KV}), which yields
\begin{gather}
\frac{\overline{V}_0}{V_0}=\frac{K_{1}+V_{1}}{K_0+V_0}, \qquad
\frac{\overline{V}_1}{V_1}=\frac{K_{0}+V_{0}}{K_1+V_1}, \nonumber\\
\frac{\overline{K}_{1}}{K_0}=\frac{1+\overline{V}_0}{1+\overline{V}_{1}}
\,\frac{\overline{K}_{1}+\overline{V}_{1}}{\overline{K}_0+\overline{V}_0},\qquad
\frac{\overline{K}_{0}}{K_1}=\frac{1+\overline{V}_1}{1+\overline{V}_{0}}
\,\frac{\overline{K}_{0}+\overline{V}_{0}}{\overline{K}_1+\overline{V}_1}.
\label{drt:1:2p}
\end{gather}
Similarly to the continuous case, we introduce additional constants
$a_n$ $(n=0,1)$ in such a way that 
\begin{gather}
\frac{\overline{V}_0}{V_0}=\frac{K_{1}+V_{1}}{K_0+V_0}\,a_0, \qquad
\frac{\overline{V}_1}{V_1}=\frac{K_{0}+V_{0}}{K_1+V_1}\,a_1, \nonumber\\
\frac{\overline{K}_{1}}{K_0}=\frac{1+\overline{V}_0}{1+\overline{V}_{1}}
\, \frac{\overline{K}_{1}+\overline{V}_{1}}{\overline{K}_0+\overline{V}_0},\qquad
\frac{\overline{K}_{0}}{K_1}=\frac{1+\overline{V}_1}{1+\overline{V}_{0}}
\, \frac{\overline{K}_{0}+\overline{V}_{0}}{\overline{K}_1+\overline{V}_1}.
\label{drt:2:2p}
\end{gather}
We remark equation~(\ref{drt:2:2p}) is still integrable in a sense that it passes
the singularity confinement test~\cite{SC}.
It is easy to see from equation~(\ref{drt:2:2p}) that we have
$\overline{K}_0\overline{K}_1=K_0K_1={\rm const}$
and $\overline{V}_0\overline{V}_1=a_0a_1V_0V_1$. We normalize as
\begin{equation}
 a_0a_1=q^{2},\qquad V_0V_1=c_0^2q^{2t}, \qquad K_0K_1=b_0^2,
\end{equation}
where $b_0$ and $c_0$ are constants. Under this normalization, we obtain
the equation,
\begin{equation}
\label{qP3:XY}
 \overline{X}YX = \frac{a_0}{q}\, \frac{1+c_0q q^{t}Y}{Y+c_0q q^t},\qquad
 YX\underline{Y} = \frac{a_0}{q}\, \frac{X+\frac{b_0}{c_0}q^{-t}}
{1+\frac{b_0}{c_0}q^{-t} X},
\end{equation}
where the variables $X=X(t)$ and $Y=Y(t)$ are defined by
\begin{equation}
 X=-\frac{a_0c_0}{b_0}q^t\,\frac{K_1}{\overline{V}_0},
\qquad Y=\frac{1}{c_0q}q^{-t}\overline{V}_0,
\end{equation}
respectively, and $\underline{\phantom{Y}}$ denotes the value at $t-1$.
Equation~(\ref{qP3:XY}) is essentially the same as $q$-P$_{\rm III}$~(\ref{qP3}). 

We finally comment on the continuous limit of $q$-P$_{\rm III}$ to P$_{\rm
III}$. The simplest way is to take the limit on the level of 
equation~(\ref{drt:2:2p}); replacing $t$, $K_i$, $V_i$ and $\alpha_i$ 
by $t/h$, $1+hK_i$, $hV_i$ and $1+\alpha_i h$ $(i=0,1)$, respectively, 
and taking the limit $h\to 0$, we obtain equation~(\ref{rt3:2p}). 

\subsection{Birational representation of affine Weyl group
 $\boldsymbol{\widetilde{W}\big(A_1^{(1)}\times A_2^{(1)}\big)}$}

In~\cite{KNY:qp4}, the following birational
representation of (extended) affine Weyl group
$\widetilde{W}\big(A_1^{(1)}\times A_2^{(1)}\big)$ is presented:

\begin{theorem}[\cite{KNY:qp4}]
Let $\boldsymbol{C}(a_i,f_i\ (i=0,1,2))$  be the field of rational functions in
 $a_i$ and $f_i$. We define the transformations $s_i$ $(i=0,1,2)$ and
 $\pi$ acting on $\boldsymbol{C}(a_i,f_i\ (i=0,1,2))$ by
\begin{gather} 
s_i(a_j)=a_j a_i^{-a_{ij}}, \qquad s_i(f_j)=f_j\left(\frac{a_i+f_i}{1+a_if_i}\right)^{u_{ij}}
\qquad (i,j=0,1,2),\nonumber\\
\pi(x_i)=x_{i+1},\qquad (x=a,f), \qquad i\in \mathbb{Z}/3\mathbb{Z},
\label{W:general}
\end{gather}
respectively, 
where $A=(a_{ij})_{i,j=0}^2$ is the generalized Cartan matrix 
of type $A^{(1)}_2$
and $U=(u_{ij})_{i,j=0}^2$ is an orientation matrix of the 
corresponding Dynkin diagram:
\begin{equation}\label{AandU}
A=\left[\begin{array}{ccc}
2 & -1 & -1 \cr
-1 & 2 & -1 \cr
-1 & -1 &  2 
\end{array}\right],\qquad
U=\left[\begin{array}{ccc}
0 & 1 & -1 \cr
-1 & 0 & 1 \cr
1 & -1 &  0
\end{array}\right]. 
\end{equation}
Moreover, we define the transformations $w_0$, $w_1$ and $r$ by
\begin{gather}
w_0(f_i)=\frac{a_ia_{i+1}(a_{i-1} a_i+a_{i-1} f_i+f_{i-1} f_i)}
{f_{i-1}(a_i a_{i+1}+a_i f_{i+1}+f_i f_{i+1})},\nonumber\\
w_1(f_i)=\frac{1+a_i f_i+a_i a_{i+1} f_i f_{i+1}}
{a_i a_{i+1} f_{i+1}(1+a_{i-1} f_{i-1}+a_{i-1} a_i f_{i-1} f_i)},\nonumber\\
r(f_i)=\frac{1}{f_i}, \qquad 
w_0(a_i)=w_1(a_i)=r(a_i)=a_i, \qquad
(i\in\mathbb{Z}/3\mathbb{Z}). \label{A11action}
\end{gather}
Then, the transformations $\langle s_0,s_1,s_2,\pi,w_0,w_1,r\rangle$ generate
the extended affine Weyl group of type $A_1^{(1)}\times A_2^{(1)}$. Namely,
these transformations satisfy the following fundamental relations,
\begin{gather}\label{affWeyl}
s_i^2=1,\qquad
(s_is_{i+1})^3=1,\qquad
\pi^3=1,\qquad
\pi s_i =s_{i+1}\pi\qquad(i=0,1,2),
\\
w_0^2=w_1^2=r^2=1, \qquad r w_0=w_1 r,
\end{gather}
and the actions of $\langle s_0,s_1,s_2,\pi\rangle$ and 
$\langle w_0,w_1,r\rangle$ commute
 with each other.
\end{theorem}

One can construct translations $T_i$ $(i=1,2,3,4)$ by
\begin{equation}
T_1=\pi s_2 s_1,\qquad T_2=s_1 \pi s_2,\qquad T_3=s_2 s_1 \pi,\qquad  T_4=r w_0.
\end{equation}
We note that these translations satisfy $T_iT_j=T_jT_i$ $(i,j=1,2,3,4)$
and $T_1T_2T_3=1$. The actions of $T_i$ on $a_i$ and $c$
are given by
\begin{align}
& T_1(a_0)=q a_0 ,&& T_1(a_1)=q^{-1}a_1,&&T_1(a_2)=a_2,&&T_1(c)=c,  \nonumber\\
& T_2(a_0)=a_0 ,&& T_2(a_1)=qa_1,&& T_2(a_2)=q^{-1}a_2,&&T_2(c)=c,\nonumber\\
& T_4(a_0)=a_0,&& T_4(a_1)=a_1,&& T_4(a_2)=a_2,&&T_4(c)=qc,
\label{Tona}
\end{align}
respectively. We also remark that $a_0a_1a_2=q$ and $f_0f_1f_2=qc^2$ are
invariant with respect to the actions of $\widetilde{W}(A_1^{(1)}\times
A_2^{(1)})$ and $\widetilde{W}(A_2^{(1)})$, respectively.

The above affine Weyl group actions determine three birational,
commutative, discrete flows on root lattice of type $A_1\times
A_2$. Regarding one of the directions as the ``time evolution'', other
directions can be considered as those of B\"acklund (Schlesinger)
transformations.

In \cite{KNY:qp4}, $T_4$ is chosen as the time evolution, whose
explicit action is given as
\begin{gather}
T_4(f_0)=a_0a_1f_1
\frac{1+a_2f_2+a_2a_0f_2f_0}{1+a_0f_0+a_0a_1f_0f_1},\qquad
T_4(f_1)=a_1a_2f_2
\frac{1+a_0f_0+a_0a_1f_0f_1}{1+a_1f_1+a_1a_2f_1f_2},\nonumber\\
T_4(f_2)=a_2a_0f_0
\frac{1+a_1f_1+a_1a_2f_1f_2}{1+a_2f_2+a_2a_0f_2f_0}.\label{T4}
\end{gather}
Choosing one of the $A_2$ direction, for example $T_1$, as the time
evolution, we have
\begin{gather}
 T_1(f_1)=f_2\frac{1+a_0f_0}{a_0+f_0}=
\frac{qc^2}{f_0f_1}\,\frac{1+a_0f_0}{a_0+f_0},\nonumber\\
 T_1^{-1}(f_0)=f_2\frac{a_1+f_1}{1+a_1f_1}=
\frac{qc^2}{f_0f_1}\,\frac{a_1+f_1}{1+a_1f_1}. \label{qP3:mapping}
\end{gather}
Operating $T_1^nT_2^\nu T_4^N$ $(n,\nu,N\in\mathbb{Z})$ on
equations~(\ref{T4}) and (\ref{qP3:mapping}), denoting $T_1^nT_2^\nu
T_4^N(f_i)=f_i(n;\nu,N)$ $(i=0,1,2)$, we obtain $q$-P$_{\rm IV}$,
\begin{gather}
\tilde f_0=a_0a_1q^{\nu}f_1\,
\frac{1+a_2q^{-\nu}f_2+a_2a_0q^{n-\nu}f_2f_0}{1+a_0q^{n} f_0+a_0a_1q^{\nu}f_0f_1},
\nonumber\\
\tilde f_1=a_1a_2q^{-n}f_2
\frac{1+a_0q^{n}f_0+a_0a_1q^{\nu}f_0f_1}{
1+a_1q^{-n+\nu}f_1+a_1a_2q^{-n}f_1f_2},\nonumber\\
\tilde f_2=a_2a_0q^{n-\nu}f_0\frac{1+a_1q^{n}f_1+a_1a_2q^{-n}f_1f_2}{
1+a_2q^{-\nu}f_2+a_2a_0q^{n-\nu}f_2f_0},\label{qP4:eq}
\end{gather}
and $q$-P$_{\rm III}$ (\ref{qP3}), respectively. In equation~(\ref{qP4:eq}),
$\tilde f_i$ denote $\tilde f_i=f_i(n;\nu,N+1)$ ($i=0,1,2$).

\begin{remark}
By construction, $q$-P$_{\rm III}$ (\ref{qP3}) describes one of the
B\"acklund transformations of $q$-P$_{\rm IV}$ (\ref{qP4:eq}),
and vice versa.  Moreover, under the specialization
(\ref{qP2:specialize}), $q$-P$_{\rm IV}$ (\ref{qP4:eq})
coincides with the B\"acklund transformation of $q$-P$_{\rm II}$
(\ref{qP2}) proposed by Joshi et al~\cite{Joshi}. 
\end{remark}

$q$-P$_{\rm III}$ admits the symmetry described by the affine Weyl group
of type $A_1^{(1)}\times A_1^{(1)}$. In fact, we can easily
verify the following fact:

\begin{proposition}
We define $w'_0$, $w'_1$ by
\begin{equation}
w'_0=s_0s_1s_0,\qquad w'_1=s_2.
\end{equation}
Then, $w'_0$ and $w'_1$ satisfy 
the fundamental relations of $W(A_1^{(1)})$,
\begin{equation}
 (w'_0)^2=(w'_1)^2=1.
\end{equation}
Moreover, the actions of $\langle w'_0,w'_1\rangle$ commute with the actions of $T_1$ and
$\langle w_0,w_1,r\rangle$. 
\end{proposition}

\section{Particular solutions of Riccati type}

\subsection{Solutions in terms of Jackson's modified $\boldsymbol{q}$-Bessel functions }

Many of Painlev\'e and discrete Painlev\'e equations are known to admit
1-parameter family of particular solutions expressible in terms of
($q$-)classical special functions. Such solutions are obtained by
finding the special cases where the Painlev\'e or discrete Painlev\'e
equations are reduced to the (discrete) Riccati equations. 

For example, $q$-P$_{\rm IV}$ (\ref{qP4:eq}) admits the following Riccati type
solutions:
\begin{proposition}[\cite{KNY:qp4}]
\label{seed:qP4}
 $q$-P$_{\rm IV}$ (\ref{qP4:eq}) admits particular solutions given by
\begin{equation}
f_0 =a_0c^2q^{n+2N+1}\frac{G_n(N)}{G_n(N+1)},\qquad 
f_1 = -\frac{q}{a_0q^{n}}\,\frac{G_n(N+1)}{G_n(N)},\qquad f_2=-1
\end{equation}
for $\nu=0$, where $G_n(N)$ is a function 
satisfying the contiguity relations
\begin{gather}
G_{n+1}(N+1)=G_{n}(N+1)+a_0^2c^2q^{2n+2N+1}G_n(N),\nonumber\\
G_{n+1}(N)=G_{n}(N+1)+c^2q^{2N+1}G_n(N).
\label{contiguity:qHW}
\end{gather}
\end{proposition}

In particular, equations (\ref{contiguity:qHW}) admit polynomial solutions in
$q^{N+1/2}c$ for $a_0=q$, and the polynomials $H_n(x)=(q^{N+1/2}c)^{-n}G_n(N)$, 
$x=(q^{N+1/2}c+q^{-N-1/2}c^{-1})/2$ are identified as {\it the 
continuous $q$-Hermite polynomials}~\cite{Gasper}. 

In the above solutions, the variables $N$ and $n$ play a role of
independent variable and parameter, respectively.  Since $q$-P$_{\rm
III}$ describes a B\"acklund transformation of $q$-P$_{\rm IV}$, the
above solutions of $q$-P$_{\rm IV}$ can be also regarded as those for
$q$-P$_{\rm III}$ only by exchanging the role of $N$ and $n$. For
notational simplicity, we introduce a parameter $\mu$ and a function~$I_\mu(n)$ by
\begin{equation}
q^{N+1/2}c = q^\mu,\qquad I_\mu(n)=(q^{N+1/2}c)^{-n} G_n(N).
\end{equation}
Then equations~(\ref{contiguity:qHW}) are rewritten as
\begin{gather}
I_{\mu+1}(n+1)=q^{-\mu}I_{\mu+1}(n)+a_0^2q^{2n+\mu}I_\mu(n),\nonumber\\
I_{\mu}(n+1)=q^{-\mu}I_{\mu+1}(n)+q^{\mu}I_\mu(n),
\label{contiguity:qmB}
\end{gather}
respectively. We note that the three-term relation of $I_\mu(n)$ 
with respect to $n$,
\begin{equation}
  I_\mu(n+2) = \left(q^{\mu}+q^{-\mu}\right)I_\mu(n+1) - \left(1-a_0^2q^{2n}\right)I_\mu(n),
\end{equation}
is nothing but Jackson's $q$-modified Bessel equation~\cite{Gasper}.

\begin{proposition}
Let $I_\mu(n)$ be a function satisfying equation~(\ref{contiguity:qmB}).
Then 
 \begin{equation}
f_0=q^{2\mu}a_0\frac{I_\mu(n)}{I_{\mu+1}(n)},\qquad 
f_1 = -\frac{q}{a_0}\, \frac{I_{\mu+1}(n)}{I_\mu(n)},
 \end{equation}
satisfy $q$-P$_{\rm III}$ (\ref{qP3}) with $\nu=0$ and $q^{N+1/2}c = q^\mu$.
\end{proposition}

We remark here that essentially the same solution is also constructed
in \cite{GNPRS,RGTT}.
By successive application of the B\"acklund transformation $T_2$ on these
solutions, we obtain ``higher-order'' solutions for
$\nu\in\mathbb{Z}_{>0}$. Moreover, such solutions admit the following
determinant formula.

\begin{theorem}\label{thm:qP4-det}
 Let $I_\mu(n)$ be a function satisfying equation~(\ref{contiguity:qmB}).
For each $\nu\in\mathbb{Z}_{\geq 0}$, we 
define a $\nu\times \nu$ determinant $\varphi_\nu^\mu(n)$ by
\begin{equation}
 \varphi_\nu^\mu(n) = \det\left(I_{\mu}(n-i+j)\right)_{1\leq i,j\leq \nu}
\quad (\nu\in\mathbb{Z}_{>0}),\qquad \varphi_0^\mu(n)=1.
\end{equation}
Then, 
\begin{equation}
  f_0=a_0q^{2\mu}\frac{\varphi_\nu^{\mu+1}(n)\varphi_{\nu+1}^{\mu}(n)}
{\varphi_{\nu+1}^{\mu+1}(n)\varphi_{\nu}^{\mu}(n)},\qquad
  f_1=-\dfrac{q}{a_0}~\dfrac{\varphi_{\nu+1}^{\mu+1}(n)\varphi_{\nu}^{\mu}(n-1)}
{\varphi_{\nu}^{\mu+1}(n-1)\varphi_{\nu+1}^{\mu}(n)},
\end{equation}
satisfy $q$-P$_{\rm III}$ (\ref{qP3}) with $\nu\in\mathbb{Z}_{\geq 0}$ 
and $q^{N+1/2}c = q^\mu$.
\end{theorem}

Since Theorem \ref{thm:qP4-det} is obtained by the interpretation of the
case of $q$-P$_{\rm IV}$~\cite{KNY:qp4}, we omit the proof. 
We also note that it is possible to construct similar determinant formula for
negative $\nu$~\cite{KNY:qp4}. 

\subsection{$\boldsymbol{q}$-P$\boldsymbol{{}_{\rm III}}$ 
specific solutions}\label{sol2}

In the previous section, we have discussed such solutions that 
$n$, $\nu$ and $N$ (or~$\mu$) played a~role of 
independent variable, determinant size and parameter of the solution,
respectively. It is possible to construct another Riccati type solution
in which the roles of $\nu$ and $N$ (or~$\mu$) are exchanged.

We first look for the special values of $N$ and $c$ which admit
decoupling of $q$-P$_{\rm III}$ (\ref{qP3}) into discrete Riccati
equation. In fact, it is easy to verify the following proposition:

\begin{proposition}
$q$-P$_{\rm III}$ (\ref{qP3}) is decoupled into the discrete Riccati equation,
\begin{equation}
 \overline{f}_1=-\frac{q^{-\nu+1}}{a_0a_1}\frac{1+a_0q^{n}f_0}{f_0},\qquad
 f_1=-a_0a_1q^{\nu}\frac{1}{a_0q^{n}+f_0},\label{ric2}
\end{equation}
when $N=0$ and $c=1$.
\end{proposition}

Eliminating $f_1$ from equation (\ref{ric2}), we obtain a discrete Riccati
equation for $f_0$,
\begin{equation}\label{ric3}
 \overline{f}_0
=\frac{-a_0q^{n+1}+\left(a_0^2a_1^2q^{2\nu-1}-a_0^2q^{2n+1}\right)f_0}
{1+a_0q^{n}f_0}.
\end{equation}
Standard linearization of equation (\ref{ric3}) gives the following result:

\begin{proposition}
Let $H_\nu(n)$ be a function satisfying the following linear difference
equation, 
 \begin{equation}
 H_\nu(n+1)=
\left[1+a_0^2a_1^2q^{2\nu} - a_0^2q^{2n}\right]H_\nu(n)-a_0^2a_1^2q^{2\nu}
H_\nu(n-1).
\end{equation}
Then, 
\begin{gather}
f_0 = \frac{1}{a_0q^{n}}\frac{H_\nu(n+1)-H_\nu(n)}{H_\nu(n)},\nonumber\\
f_1 = -a_0^2a_1q^{n+\nu}
\frac{H_\nu(n)}{H_\nu(n+1)-\left(1-a_0^2q^{2n}\right)H_\nu(n)},
\end{gather}
satisfy $q$-P$_{\rm III}$ (\ref{qP3}) for $N=0$ and $c=1$.
\end{proposition}

One can investigate the evolution with respect to $\nu$ by substituting the
above expression into the B\"acklund transformations of $q$-P$_{\rm
III}$ described in Section 2.2. After some manipulation we arrive at the
following result:

\begin{theorem}
Let  $H_\nu(n)$ be a function satisfying the following contiguity relations,
\begin{gather}
H_\nu(n+1) - H_\nu(n) = -a_0^2q^{2n}H_{\nu-1}(n),\label{H:rec1}\\ 
H_{\nu+1}(n) - H_\nu(n) = -a_0^2a_1^2q^{2\nu}H_{\nu}(n-1).\label{H:rec2}
\end{gather}
Then,
\begin{equation}
f_0 = -a_0q^{n}\frac{H_{\nu-1}(n)}{H_\nu(n)},\qquad
f_1 = \frac{1}{a_0^2a_1q^{n+\nu-2}}\frac{H_\nu(n)}{H_{\nu-1}(n-1)},
\end{equation}
satisfy $q$-P$_{\rm III}$ (\ref{qP3})  with $N=0$ and $c=1$.
\end{theorem}

\begin{remark}
As is expected from the symmetry described in Section 2.2, the
contiguity relations (\ref{H:rec1}) and (\ref{H:rec2}) for $H_\nu(n)$
are symmetric with respect to $n$ and $\nu$. In particular, the
three-term relation with respect to $\nu$ is obtained as
\begin{equation}
 H_{\nu+1}(n)=\left[1 + a_0^2q^{2n} - a_0^2a_1^2q^{2\nu} \right]H_{\nu}(n)-
a_0^2q^{2n} H_{\nu-1}(n).
\end{equation} 
\end{remark}

We obtain higher-order solutions in the form of rational
functions in $H_\nu(n)$ by applying the B\"acklund
transformation $T_4$ to the above solution. Furthermore, these solutions admit the
following determinant formula:

\begin{theorem}\label{qP3:determinant formula}
For each $N\in\mathbb{Z}_{\geq 0}$, we define an $N\times N$ determinant
$\phi_\nu^N(n)$ by
\begin{equation}
\phi_\nu^N(n)=
\det \left(H_{\nu+i-1}(n+j-1) \right)_{1\leq i,j\leq N}
\quad (N\in\mathbb{Z}_{>0}),\qquad  \phi_\nu^0(n)=1.\label{phi}
\end{equation}
Then,
\begin{gather}
 f_0=-a_0q^{n+2N}~\frac{\phi_{\nu-1}^{N+1}(n)\phi_\nu^N(n)}
{\phi_\nu^{N+1}(n)\phi_{\nu-1}^N(n)}, \nonumber\\
 f_1=\frac{1}{a_0^2a_1q^{n+\nu+2N-2}}
\frac{\phi_\nu^{N+1}(n)\phi_{\nu-1}^N(n-1)}
{\phi_{\nu-1}^{N+1}(n-1)\phi_\nu^N(n)},
\label{f}
\end{gather}
satisfy $q$-P$_{\rm III}$ (\ref{qP3}) with $c=1$, $N\in \mathbb{Z}_{\geq 0}$.
\end{theorem}

We next prove Theorem \ref{qP3:determinant formula}. It is a direct
consequence of the following ``multiplicative formula'' with respect to $\phi$.

\begin{proposition}\label{multi}
The following formulas holds.
 \begin{gather}
 1+a_0q^nf_0 = q^{-2N}
\frac{\phi_\nu^{N+1}(n+1)\phi_{\nu-1}^N(n-1)}
{\phi_\nu^{N+1}(n)\phi_{\nu-1}^N(n)},\label{mul1}\\
 1+\frac{f_0}{a_0q^n}= -a_0^2a_1^2q^{2(\nu+N-1)}
\frac{\phi_{\nu-1}^{N+1}(n-1)\phi_\nu^N(n+1)}
{\phi_\nu^{N+1}(n)\phi_{\nu-1}^N(n)},\label{mul2}\\
 1+a_1q^{-n+\nu}f_1 = \frac{1}{a_0^2q^{2(n-1)}}
\frac{\phi_\nu^{N+1}(n-1)\phi_{\nu-1}^N(n)}
{\phi_{\nu-1}^{N+1}(n-1)\phi_\nu^N(n)},\label{mul3}\\
 1+\frac{f_1}{a_1q^{-n+\nu}} = \frac{1}{a_0^2a_1^2q^{2(\nu-1)}}
\frac{\phi_{\nu-1}^{N+1}(n)\phi_\nu^N(n-1)}
{\phi_{\nu-1}^{N+1}(n-1)\phi_\nu^N(n)}.\label{mul4}
\end{gather}
\end{proposition}

In fact, it is an easy task to derive $q$-P$_{\rm III}$ by using the
above multiplicative formulas and equation (\ref{f}).
Furthermore, Proposition~\ref{multi} follows from the following
proposition.

\begin{proposition}\label{bilinear}
 $\phi_\nu^N(n)$ satisfies the following bilinear difference equations.
\begin{gather}
\phi_\nu^{N+1}(n)\phi_{\nu-1}^N(n)
- a_0^2q^{2(n+N)} \phi_{\nu-1}^{N+1}(n)\phi_\nu^N(n)
\nonumber\\
\qquad \qquad {}= q^{-2N}\phi_{\nu}^{N+1}(n+1)\phi_{\nu-1}^N(n-1),\label{bl1}\\
q^{2N}\phi_{\nu-1}^{N+1}(n)\phi_\nu^N(n)
- a_0^2a_1^2q^{2(\nu+N-1)} \phi_{\nu-1}^{N+1}(n-1)\phi_\nu^N(n+1)
\nonumber\\
\qquad \qquad {} = \phi_\nu^{N+1}(n)\phi_{\nu-1}^N(n),\label{bl2}\\
a_0^2q^{2(n-1)}\phi_{\nu-1}^{N+1}(n-1)\phi_\nu^N(n)
- q^{-2N} \phi_\nu^{N+1}(n)\phi_{\nu-1}^N(n-1)\nonumber\\
\qquad \qquad {} = \phi_\nu^{N+1}(n-1)\phi_{\nu-1}^N(n),\label{bl3}\\
q^{2(\nu-1)}\phi_{\nu-1}^{N+1}(n-1)\phi_\nu^N(n)
- q^{-2N} \phi_\nu^{N+1}(n)\phi_{\nu-1} ^N(n-1)\nonumber\\
\qquad \qquad {}= \phi_{\nu-1}^{N+1}(n)\phi_\nu^N(n-1).\label{bl4}
\end{gather}
\end{proposition}

For example, equation (\ref{mul1}) is derived from equation (\ref{bl1}) by dividing
both sides by $\phi_\nu^{N+1}(n)\phi_{\nu-1}^N(n)$. Therefore, the proof
of Theorem~\ref{qP3:determinant formula} is now reduced to that of
Proposition~\ref{bilinear}. We will give it in the next section.

\subsection{Proof of Proposition \ref{bilinear}}

Our basic idea for proving Proposition~\ref{bilinear} is the same as 
that was done in \cite{KNY:qp4}.
Bilinear difference equations are derived from the Pl\"ucker
relations, which are quadratic identities among determinants whose
columns are shifted. Therefore, we first construct such ``difference
formulas'' that relate ``shifted determinants'' and $\phi_{\nu}^N(n)$, by
using the contiguity relations of $H_\nu(n)$. We then derive bilinear
difference equations with the aid of difference formulas from proper
Pl\"ucker relations.  We take equation (\ref{bl2}) as an example to show this
procedure explicitly.  For other equations, see Appendix~A.

We introduce a notation,
\begin{gather}
 \phi_\nu^N(n)=
\left|
\begin{array}{cccc}
 H_\nu(n)& H_{\nu+1}(n)&\cdots &H_{\nu+N-1}(n) \\
 H_{\nu}(n+1)& H_{\nu+1}(n+1)&\cdots &H_{\nu+N-1}(n+1) \\
\vdots & \vdots & \ddots & \vdots\\
 H_{\nu}(n+N-1)& H_{\nu+1}(n+N-1)&\cdots &H_{\nu+N-1}(n+N-1)
\end{array}
\right|\nonumber\\
\phantom{\phi_\nu^N(n)}{}=|\boldsymbol{0}_n,\boldsymbol{1}_n,\cdots,
\boldsymbol{N}-\boldsymbol{1}_n|,\label{vec}
\end{gather}
where $\boldsymbol{k}_n$ denotes a column vector,
\begin{equation}
 \boldsymbol{k}_n=\left(\begin{array}{c}
             H_{\nu+k}(n)\\H_{\nu+k}(n+1)\\ \vdots
                  \end{array}\right).
\end{equation}
We note that we abbreviate the suffix $n$ when unnecessary.

We next construct the difference formula.
\begin{lemma}[Difference formula I] \label{Difference Formula I}
 \begin{gather}
|\boldsymbol{0},\boldsymbol{1},\cdots,\boldsymbol{N}-\boldsymbol{1}|
= \phi_\nu^N(n),\label{difI1}\\
|\boldsymbol{0}_{n+1},\boldsymbol{0},\cdots,\boldsymbol{N}-\boldsymbol{2}|
=\left(-a_0^2a_1^2\right)^{-(N-2)}q^{-(2\nu+N-2)(N-1)} \phi_\nu^N(n+1),\label{difI2}\\
|\boldsymbol{1}_{n+1},\boldsymbol{0},\cdots,\boldsymbol{N}-\boldsymbol{2}|
=\left(-a_0^2a_1^2\right)^{-(N-2)}q^{-(2\nu+N-2)(N-1)} \phi_\nu^N(n+1).\label{difI3}
\end{gather}
\end{lemma}


\begin{proof}
Equation (\ref{difI1}) is nothing but the definition equation (\ref{vec}).
Subtracting the $(k-1)$-th
column from the $k$-th column of the left hand side of equation~(\ref{difI1})
and using the contiguity relation~(\ref{H:rec2}) for
$k=N,N-1,\ldots,2$, we have
\begin{gather}
  \phi_\nu^N(n)=\left|
\begin{array}{ccc}
 \cdots &H_{\nu+N-2}(n)&H_{\nu+N-1}(n)-H_{\nu+N-2}(n) \\
\cdots &H_{\nu+N-2}(n+1)&
H_{\nu+N-1}(n+1)-H_{\nu+N-2}(n+1)\\
 \vdots & \vdots & \vdots\\
\cdots &H_{\nu+N-2}(n+N-1)&H_{\nu+N-1}(n+N-1) 
-H_{\nu+N-2}(n+N-1)
\end{array}
\right|\!\!\nonumber
\end{gather}
\begin{gather}
\phantom{\phi_\nu^N(n)}{}=\left|
\begin{array}{ccc}
\cdots &H_{\nu+N-2}(n)&-a_0^2a_1^2q^{2(\nu+N-2)}H_{\nu+N-2}(n-1) \\
\cdots &H_{\nu+N-2}(n+1)&-a_0^2a_1^2q^{2(\nu+N-2)}H_{\nu+N-2}(n) \\
\vdots & \vdots & \vdots\\
\cdots &H_{\nu+N-2}(n+N-1)
&-a_0^2a_1^2q^{2(\nu+N-2)}H_{\nu+N-2}(n+N-2)
\end{array}
\right|\nonumber\\
\phantom{\phi_\nu^N(n)}{} = -a_0^2a_1^2q^{2(\nu+N-2)}\nonumber\\
\phantom{\phi_\nu^N(n)=}{}\times\left|
\begin{array}{cccc}
 H_\nu(n)&\cdots &H_{\nu+N-2}(n)&H_{\nu+N-2}(n-1) \\
 H_{\nu}(n+1)&\cdots &H_{\nu+N-2}(n+1)&H_{\nu+N-2}(n) \\
\vdots &\vdots & \vdots & \vdots\\
 H_{\nu}(n+N-1)&\cdots &H_{\nu+N-2}(n+N-1)&H_{\nu+N-2}(n+N-2)
\end{array}
\right|\nonumber\\
\phantom{\phi_\nu^N(n)}{} =\cdots\cdots\cdots\cdots\cdots\cdots \nonumber\\
\phantom{\phi_\nu^N(n)}{}= \left(-a_0^2a_1^2\right)^{N-1}q^{(2\nu+N-2)(N-1)}\nonumber\\
\phantom{\phi_\nu^N(n)=}{}\times
\left|
\begin{array}{cccc}
 H_\nu(n)      & H_{\nu}(n-1)&\cdots &H_{\nu+N-2}(n-1) \\
 H_{\nu}(n+1)& H_{\nu}(n)  &\cdots &H_{\nu+N-2}(n) \\
\vdots &\vdots & \vdots & \vdots\\
 H_{\nu}(n+N-1)& H_{\nu}(n+N-2)&\cdots &H_{\nu+N-2}(n+N-2)
\end{array}
\right|,\label{difI22}
\end{gather}
which is nothing but equation (\ref{difI2}). Moreover, subtracting the second
column multiplied by $a_0^2a_1^2q^{2\nu}$ 
from the first column of the right hand side of equation (\ref{difI22}) and
using the contiguity relation (\ref{H:rec2}), we obtain
\begin{gather}
  \phi_\nu^N(n)=\left(-a_0^2a_1^2\right)^{N-1}q^{(2\nu+N-2)(N-1)}\nonumber\\
\qquad{}\times
\left|
\begin{array}{cccc}
 H_{\nu+1}(n)      & H_{\nu}(n-1)&\cdots &H_{\nu+N-2}(n-1) \\
 H_{\nu+1}(n+1)& H_{\nu}(n)  &\cdots &H_{\nu+N-2}(n) \\
\vdots &\vdots & \vdots & \vdots\\
 H_{\nu+1}(n+N-1)& H_{\nu}(n+N-2)&\cdots &H_{\nu+N-2}(n+N-2)
\end{array}
\right|,
\end{gather}
which is nothing but equation~(\ref{difI3}).
\end{proof}

Now consider the  Pl\"ucker relation,
\begin{gather}
 0=
|\boldsymbol{1}_{n+1},\boldsymbol{1},\cdots,\boldsymbol{N}-\boldsymbol{1},\boldsymbol{N}|
\times
|\boldsymbol{0},\boldsymbol{1},\cdots,\boldsymbol{N}-\boldsymbol{1},\varphi_2|\nonumber\\
\phantom{0=}{}-|\boldsymbol{0},\boldsymbol{1},\cdots,\boldsymbol{N}-\boldsymbol{1},
\boldsymbol{N}|
\times
|\boldsymbol{1}_{n+1},\boldsymbol{1},\boldsymbol{1},\cdots,\boldsymbol{N}-\boldsymbol{1},
\varphi_2|\nonumber\\
\phantom{0=}{}-|\boldsymbol{1},\cdots,\boldsymbol{N}-\boldsymbol{1},\boldsymbol{N},\varphi_2|
\times
|\boldsymbol{1}_{n+1},\boldsymbol{0},\boldsymbol{1},\cdots,\boldsymbol{N}-
\boldsymbol{1}|,\label{pl1}
\end{gather}
where $\varphi_2$ is the column vector,
\begin{equation}
 \varphi_2 = \left(\begin{array}{c}1 \\0\\\vdots\\0
                   \end{array}\right).
\end{equation}
We obtain equation~(\ref{bl2}) by applying Lemma~\ref{Difference Formula I} to
equation~(\ref{pl1}) after expanding with respect to the column~$\varphi_2$.
This completes the proof of bilinear equation~(\ref{bl2}).

\appendix

\section{Difference formulas and Pl\"ucker relations}

In this appendix, we provide with data which are necessary for the proof of
Proposition~\ref{bilinear}. We first give the difference formulas.

\begin{lemma}[Difference formula II]\label{Difference Formula II}
\begin{gather}
|\boldsymbol{0}',\boldsymbol{1}',\cdots,\boldsymbol{N}-\boldsymbol{1}'|= \phi_{\nu}^N(n),\\
|\boldsymbol{0}'_{\nu+1},\boldsymbol{0}',\cdots,\boldsymbol{N}-\boldsymbol{2}'|
=\left(-a_0^2\right)^{-(N-2)}q^{-(2n+N-2)(N-1)} \phi_{\nu+1}^N(n),\\
|\boldsymbol{1}'_{\nu+1},\boldsymbol{0}',\cdots,\boldsymbol{N-2}'|
=\left(-a_0^2\right)^{-(N-2)}q^{-(2n+N-2)(N-1)} \phi_{\nu+1}^N(n),
\end{gather}
where
\begin{equation}
 \boldsymbol{k}'_\nu=\left(\begin{array}{c}
             H_{\nu}(n+k)\\H_{\nu+1}(n+k)\\\vdots
                  \end{array}\right).
\end{equation}
\end{lemma}

We next rewrite $\phi_\nu^N(n)$ in terms of Toeplitz determinant with
respect to $n$ by using the contiguity relations (\ref{H:rec1}) and~(\ref{H:rec2}) as
\begin{gather}
 \phi_{\nu}^N(n) = \left(-a_0^2a_1^2\right)^{N(N-1)/2}q^{N(N-1)(3\nu+N-2)/3}\nonumber\\
\phantom{\phi_{\nu}^N(n) =}{}\times
\left|
\begin{array}{cccc}
 H_{\nu}(n)&H_{\nu}(n+1) &\cdots &H_{\nu}(n+N-1) \\
 H_{\nu}(n-1)&H_{\nu}(n) &\cdots &H_{\nu}(n+N-2) \\
\vdots & \vdots & \ddots & \vdots\\
 H_{\nu}(n-N+1)&H_{\nu}(n-N+2) &\cdots &H_{\nu}(n)
\end{array}
\right|\nonumber\\
\phantom{\phi_{\nu}^N(n)}{}\equiv
\left(-a_0^2a_1^2\right)^{N(N-1)/2}q^{N(N-1)(3\nu+N-2)/3}~\psi_\nu^N(n).\label{psi}
\end{gather}

Then we obtain the third difference formula:
\begin{lemma}[Difference formula III]\label{Difference Formula III}
\begin{gather}
|\overline{\boldsymbol{0}},\overline{\boldsymbol{1}},\cdots,
\overline{\boldsymbol{N}-\boldsymbol{1}}|
= \psi_{\nu}^N(n),\\
|\widetilde{\overline{\boldsymbol{0}}}_{\nu+1},\overline{\boldsymbol{0}},\cdots,
\overline{\boldsymbol{N}-\boldsymbol{2}}|
=\left(-a_0^2\right)^{-(N-2)}q^{-2(n-1)(N-1)} \psi_{\nu+1}^N(n),\\
|\widetilde{\overline{\boldsymbol{1}}}_{\nu+1},\overline{\boldsymbol{0}},
\cdots,\overline{\boldsymbol{N}-\boldsymbol{2}}|
=\left(-a_0^2\right)^{-(N-2)}q^{-2(n-1)(N-1)} \psi_{\nu+1}^N(n),
\end{gather}
where
\begin{equation}
 \overline{\boldsymbol{k}}_\nu=\left(\begin{array}{c}
             H_{\nu}(n+k)\\H_{\nu}(n+k-1)\\H_\nu(n+k-2)\\\vdots
                  \end{array}\right),
\qquad
 \widetilde{\overline{\boldsymbol{k}}}_\nu=\left(\begin{array}{c}
             H_{\nu}(n+k)\\q^{-2}H_{\nu}(n+k-1)\\
q^{-4}H_{\nu}(n+k-2)\\\vdots
                  \end{array}\right).
\end{equation} 
\end{lemma}

Lemmas \ref{Difference Formula II} and \ref{Difference Formula III} are
proved by applying the same procedure as that for 
Lemma~\ref{Difference Formula I} on the determinants,
\begin{equation}
 \phi_\nu^N(n)=\left|
\begin{array}{cccc}
 H_\nu(n)& H_\nu(n+1)&\cdots &H_\nu(n+N-1) \\
 H_{\nu+1}(n)& H_{\nu+1}(n+1)&\cdots &H_{\nu+1}(n+N-1) \\
\vdots & \vdots & \ddots & \vdots\\
 H_{\nu+N-1}(n)& H_{\nu+N-1}(n+1)&\cdots &H_{\nu+N-1}(n+N-1)
\end{array}
\right|,
\end{equation}
and
\begin{equation}
\psi_\nu^N(n) =  
\left|
\begin{array}{cccc}
 H_{\nu}(n)&H_{\nu}(n+1) &\cdots &H_{\nu}(n+N-1) \\
 H_{\nu}(n-1)&H_{\nu}(n) &\cdots &H_{\nu}(n+N-2) \\
\vdots & \vdots & \ddots & \vdots\\
 H_{\nu}(n-N+1)&H_{\nu}(n-N+2) &\cdots &H_{\nu}(n)
\end{array}
\right|,
\end{equation}
respectively, by using the contiguity relation (\ref{H:rec1}).

The bilinear equations (\ref{bl1})--(\ref{bl4}) are derived by
considering proper Pl\"ucker relations. We finally give the list the
Pl\"ucker relations and difference formulas which are necessary for the
derivations. This completes the proof of Proposition~\ref{bilinear} and
thus Theorem~\ref{qP3:determinant formula}.

\medskip

\noindent \textbf{Equation (\ref{bl1}):}\\
{\bf Pl\"ucker relation}
\begin{gather*}
 0= |\widetilde{\overline{\boldsymbol{1}}}_{\nu+1},\overline{\boldsymbol{1}},\cdots,
\overline{\boldsymbol{N}-\boldsymbol{1}},\overline{\boldsymbol{N}}|
\times
|\overline{\boldsymbol{0}},\overline{\boldsymbol{1}},\cdots,
\overline{\boldsymbol{N}-\boldsymbol{1}},\varphi_2|\\
\phantom{0=}{}-
|\overline{\boldsymbol{0}},\overline{\boldsymbol{1}},\cdots,
\overline{\boldsymbol{N}-\boldsymbol{1}},\overline{\boldsymbol{N}}|
\times
|\widetilde{\overline{\boldsymbol{1}}}_{\nu+1},\overline{\boldsymbol{1}},\cdots,
\overline{\boldsymbol{N}-\boldsymbol{1}},\varphi_2|\\
\phantom{0=}{}-
|\overline{\boldsymbol{1}},\cdots,
\overline{\boldsymbol{N}-\boldsymbol{1}},\overline{\boldsymbol{N}},\varphi_2|
\times
|\widetilde{\overline{\boldsymbol{1}}}_{\nu+1},
\overline{\boldsymbol{0}},\overline{\boldsymbol{1}},\cdots,
\overline{\boldsymbol{N}-\boldsymbol{1}}|.
\end{gather*}
{\bf Difference formula} Lemma \ref{Difference Formula III}.

\medskip

\noindent \textbf{Equation (\ref{bl3}):}\\
{\bf Pl\"ucker relation}
\begin{gather*}
 0=|\boldsymbol{1}'_{\nu+1},\boldsymbol{1}',\cdots,\boldsymbol{N}-\boldsymbol{1}',
\boldsymbol{N}'|
\times
|\boldsymbol{0}',\boldsymbol{1}',\cdots,\boldsymbol{N}-
\boldsymbol{1}',\varphi_1|\\
\phantom{0=}{}-|\boldsymbol{0}',\boldsymbol{1}',\cdots,\boldsymbol{N}
-\boldsymbol{1}',\boldsymbol{N}'|
\times
|\boldsymbol{1}'_{\nu+1},\boldsymbol{1}',\boldsymbol{1}',\cdots,\boldsymbol{N}-
\boldsymbol{1}',\varphi_1|\\
\phantom{0=}{}-|\boldsymbol{1}',\cdots,\boldsymbol{N}-
\boldsymbol{1}',\boldsymbol{N}',\varphi_1|
\times
|\boldsymbol{1}'_{\nu+1},\boldsymbol{0}',\boldsymbol{1}',\cdots,\boldsymbol{N}-
\boldsymbol{1}'|.
\end{gather*}
{\bf Difference formula} Lemma \ref{Difference Formula II}.

\medskip

\noindent \textbf{Equation (\ref{bl4}):}\\
{\bf Pl\"ucker relation}
\begin{gather*}
 0= |\boldsymbol{1}_{n+1},\boldsymbol{1},\cdots,\boldsymbol{N}-
\boldsymbol{1},\boldsymbol{N}|
\times
|\boldsymbol{0},\boldsymbol{1},\cdots,\boldsymbol{N}-
\boldsymbol{1},\varphi_1|\\
\phantom{0=}{}-|\boldsymbol{0},\boldsymbol{1},\cdots,\boldsymbol{N}-
\boldsymbol{1},\boldsymbol{N}|
\times
|\boldsymbol{1}_{n+1},\boldsymbol{1},\boldsymbol{1},\cdots,\boldsymbol{N}-
\boldsymbol{1},\varphi_1|\\
\phantom{0=}{}-|\boldsymbol{1},\cdots,\boldsymbol{N}-
\boldsymbol{1},\boldsymbol{N},\varphi_1|
\times
|\boldsymbol{1}_{n+1},\boldsymbol{0},\boldsymbol{1},\cdots,\boldsymbol{N}-
\boldsymbol{1}|.
\end{gather*}
{\bf Difference formula} Lemma \ref{Difference Formula I}.

\medskip

In the above data, $\varphi_1$ and $\varphi_2$ are column vectors given by
\begin{equation}
 \varphi_1=\left(\begin{array}{c}
            0\\\vdots\\0\\1
                 \end{array}\right),\qquad
 \varphi_2=\left(\begin{array}{c}
            1\\0\\\vdots\\0
                 \end{array}\right).
\end{equation}

\label{kajiwara-lastpage}

\end{document}